# Confinement-Induced Chiral Edge Channel Interaction in Quantum Anomalous Hall Insulators

Ling-Jie Zhou[1,3], Ruobing Mei[1,3], Yi-Fan Zhao[1,3], Ruoxi Zhang[1], Deyi Zhuo[1], Zi-Jie Yan[1], Wei Yuan[1], Morteza Kayyalha[1,2], Moses H. W. Chan[1], Chao-Xing Liu[1], and Cui-Zu Chang[1]

[1] Department of Physics, The Pennsylvania State University, University Park, PA 16802, USA

[2] Department of Electrical Engineering, The Pennsylvania State University, University Park, Pennsylvania 16802, USA

[3] These authors contributed equally: Ling-Jie Zhou, Ruobing Mei, Yi-Fan Zhao

Corresponding authors: cxl56@psu.edu (C.-X. L.); cxc955@psu.edu (C.-Z. C.)

**Abstract:** In quantum anomalous Hall (QAH) insulators, the interior is insulating but electrons can travel with zero resistance along one-dimensional conducting paths known as chiral edge channels (CECs). These CECs have been predicted to be confined to the one-dimensional (1D) edges and exponentially decay in the two-dimensional (2D) bulk. In this work, we present the results of a systematic study of QAH devices fashioned in a Hall bar geometry of different widths. At the charge neutral point, the QAH effect persists in a Hall bar device with a width of only ~72 nm, implying the intrinsic decaying length of CECs is less than ~36 nm. In the electron-doped regime, we find that the Hall resistance deviates quickly from the quantized value when the sample width is less than 1 $\mu$m. Our theoretical calculations suggest that the deviation from the quantized Hall resistance in narrow QAH samples originates from the interaction between two opposite CECs mediated by disorder-induced bulk states in QAH insulators, consistent with our experimental observations.

**Main text:** Current flowing through a semiconductor inevitably incurs energy dissipation, the management of which is a key challenge for circuit miniaturization. The search for alternative paradigms of electronics is a vibrant frontier in condensed matter physics. Two solid-state phenomena exhibit resistance-free current. One is superconductivity, which usually arises in metallic materials with high carrier density and is thus difficult to be electrically manipulated. The other one is the chiral edge channel (CEC) formed in quantum Hall insulators [1]. The CEC can be easily switched on/off by applying a gate voltage, but the need for a high external magnetic field hampers its potential applications in electronic devices. The necessity of an external magnetic field is circumvented in the quantum anomalous Hall (QAH) effect realized in the magnetically doped topological insulator (TI) films/heterostructures [2-10]. Like the quantum Hall insulators, the QAH insulators also possess the dissipation-free CEC. Therefore, the QAH insulator is an outstanding quantum coherent platform for next-generation electronics and spintronics as well as topological quantum computations [2,11-15].

The QAH effect is first realized experimentally in a TI film doped with transition metal ions using molecular beam epitaxy (MBE), specifically 5 quintuple layers (QL) Cr-doped $(Bi, Sb)_2Te_3$ [3]. Soon after, the QAH effect is also realized in a 4 QL V-doped $(Bi, Sb)_2Te_3$ film [4,5]. To date, the QAH effect is usually realized in mechanically scratched millimeter-size [2-7,9,10,16-18] and photolithography-patterned tens to hundreds of micrometer-size [8,19-24] Hall bar devices. In these QAH devices, two CECs on opposite sides of the sample are well separated (Fig. 1a) and thus the QAH effect usually exhibits quantized Hall resistance and zero longitudinal resistance. However, to access the potential applications, the miniaturized QAH devices within the quantum coherent length are essential. Moreover, prior theoretical studies predicted when the width of the QAH Hall bar $w$ is less than $2d_0$ ($d_0$ is the CEC width), the two

CECs couple to each other and thus the QAH effect disappears [25] (Fig. 1a). By employing the combining photolithography and electron-beam lithography fabrication techniques, the QAH effect is recently observed in magnetic TI Hall bar devices with $w$ of ~600 nm [26] and ~160 nm [27]. However, studies on how the QAH state evolves with width in Hall bar devices down to tens of nanometers are still absent. Moreover, the helical-like conducting channels in quasi-1D QAH structures with $w < 2d_0$ are expected to possess a broad topological regime for the formation of the localized Majorana zero modes by introducing a superconducting order through the proximity effect [25,28]. Therefore, patterning the QAH device with a narrow width that approaches $2d_0$ is crucial for the development of scalable topological quantum computations [2].

In this work, we fabricate QAH Hall bar devices with widths that range from 72 nm to 100 μm from molecular beam epitaxy (MBE)-grown magnetic TI/TI sandwich heterostructures, specifically 3 QL Cr-doped $(Bi,Sb)_2Te_3$/4 QL $(Bi,Sb)_2Te_3$/3 QL Cr-doped $(Bi,Sb)_2Te_3$ [10,29]. We perform transport measurements to study the confinement-induced interaction between two CECs in QAH insulators. At the charge neutral point, we find that the QAH state persists in a Hall bar device with a width of only ~72 nm, so the width of the CEC $d_0$ is less than ~36 nm, which is much smaller than the value revealed in prior microscopy studies on similar QAH samples [21,30]. However, in the electron-doped regime, we find that the Hall resistance deviates quickly from the quantized value when $w$ is less than 1 μm. We theoretically model the decaying behaviors of CECs away from the edges and find that the density of states of CECs comprises an exponential decay part with a short decay length and a bulk-like part with a much longer localization length. Our experimental and theoretical studies suggest the derivation from the quantized Hall resistance in narrow QAH samples originates from the interaction between two opposite CECs mediated by disorder-induced bulk states in QAH insulators.

All QAH sandwich samples are grown on heat-treated SrTiO$_3$(111) substrates in an MBE system (Omicron Lab 10) with a vacuum better than ~ $2\times10^{-10}$ mbar. The QAH Hall bar devices with $w$ from 72 nm to 10 µm are fabricated using electron-beam lithography, while the ones with $w$ between 10 µm and 100 µm are fabricated using photolithography [31]. A bottom gate voltage $V_g$ is employed to tune the chemical potential of the QAH devices. A scanning electron microscope image of the ~72 nm QAH Hall bar device is shown in Fig. 1b. The transport measurements are carried out by the standard lock-in amplifier technique with ~1 nA excitation current in a dilution refrigerator (Leiden Cryogenics, 10 mK, 9 T). More details of the MBE growth, device fabrication, and transport measurements can be found in Supplemental Material [31].

We first focus on the ~72 nm QAH Hall bar device. Figure 1c shows the magnetic field $\mu_0 H$ dependence of the longitudinal resistance $\rho_{xx}$ and the Hall resistance $\rho_{yx}$ measured at the charge neutral point $V_g = V_g^0$ and $T = 25$ mK. The value of $\rho_{yx}$ at zero magnetic field [labeled as $\rho_{yx}(0)$] is found to be ~0.9517 $h/e^2$ and $\rho_{xx}(0)$ is ~0.1911 $h/e^2$. The ratio $\rho_{yx}(0)/\rho_{xx}(0)$ corresponds to an anomalous Hall angle $\alpha$~78.65°, indicating the chiral edge transport still dominates over the bulk transport, i.e. the QAH state persists in this ~72 nm Hall bar device [2]. The non-zero $\rho_{xx}(0)$ is expected to be caused by the confinement effect-induced interaction between two opposite CECs in quasi-1D QAH insulators (Fig. 1a), which we will discuss in detail below. We note that there are fluctuations observed in $\rho_{xx}$ and $\rho_{yx}$, particularly near the coercive field $\mu_0 H_c$ regime. The fluctuations are presumably a result of the fact that the QAH bar width $w$ is comparable with the magnetic domain size [26,27,32,33]. The QAH state in the ~72 nm Hall bar device is further demonstrated by the gate ($V_g$-$V_g^0$) dependence of $\rho_{yx}(0)$ and $\rho_{xx}(0)$ (Fig. 1d). The nearly quantized $\rho_{yx}(0)$ peak and the sharp $\rho_{xx}(0)$ dip near $V_g = V_g^0$ validate the existence of the QAH state.

The evolution of the QAH state is demonstrated in the transport results of QAH Hall bars with $w$ from 100 μm down to 72 nm. Figures 2a to 2f show the $\mu_0 H$ dependence of $\rho_{xx}$ and $\rho_{yx}$ of the QAH Hall bar devices with 300 nm ≤ $w$ ≤ 10 μm measured at $V_g = V_g^0$ and $T$ =25 mK. For the ~300 nm and ~500 nm QAH Hall bar devices, $\rho_{yx}(0)$ shows a nearly quantized value of ~0.9545 $h/e^2$ and ~0.9772 $h/e^2$ at $V_g = V_g^0$, concomitant with $\rho_{xx}(0)$ ~0.1270 $h/e^2$ and ~0.0275 $h/e^2$, respectively (Figs. 2a and 2b). We find that the QAH state is steadily improved with increasing $w$. For the 1 μm ≤ $w$ ≤ 10 μm QAH Hall bar devices, $\rho_{yx}(0)$ shows a quantized value of ~0.9854 $h/e^2$, ~0.9864 $h/e^2$, ~0.9880 $h/e^2$, and ~0.9871 $h/e^2$ for the $w$ ~1 μm, 2 μm, 5 μm, and 10 μm samples, respectively. The corresponding $\rho_{xx}(0)$ is ~0.0107 $h/e^2$, ~0.0121 $h/e^2$, ~0.0093 $h/e^2$, and ~0.0035 $h/e^2$, respectively (Figs. 2c to 2d). The Hall bar devices with 10 μm ≤ $w$ ≤100 μm and $w$ = 500 μm exhibit the perfect QAH effect (Figs. S2 to S4). These systematic measurements show that as soon as $w$ is reduced to less than ~1 μm, the $\rho_{xx}(0)$ value begins to increase while the $\rho_{yx}(0)$ value slightly drops. This behavior illustrates the interaction between two CECs starts occurring in our QAH devices with $w$ ~1 μm.

The QAH states in these devices are further validated by the gate ($V_g$-$V_g^0$) dependence of $\rho_{yx}(0)$ and $\rho_{xx}(0)$ (Figs. 3, S3, and S4). For each QAH sample, $\rho_{yx}(0)$ exhibits a peak or plateau, demonstrating the existence of the QAH state at $V_g = V_g^0$. Moreover, we observe substantially different decaying behaviors of the QAH states for $V_g = V_g^0$ and $V_g > V_g^0$. For $V_g = V_g^0$, $\rho_{yx}(0)$ deviates slowly from $h/e^2$, and $\rho_{xx}(0)$ progressively increases for $w$ ≤1 μm (Figs. 3a to 3f). The ratios $\rho_{yx}(0)/\rho_{xx}(0)$ correspond to anomalous Hall angles of ~82.4225°, ~88.3880°, ~89.3802°,

~89.2972°, ~89.4594°, and ~89.7986° for the $w$ ~300 nm, 500 nm, 1 μm, 2 μm, 5 μm, and 10 μm samples, respectively (Figs. 3a to 3f). For $V_g < V_g^0$, these QAH devices show similar gate ($V_g$-$V_g^0$) dependence of $\rho_{yx}(0)$ and $\rho_{xx}(0)$ behaviors for all QAH devices, deviating from the QAH state very quickly as $V_g$ is tuned away from $V_g^0$. However, for $V_g > V_g^0$, the gate ($V_g$-$V_g^0$) dependence of $\rho_{yx}(0)$ and $\rho_{xx}(0)$ behaviors with different $w$ are substantially different and show a systematic trend. By reducing $w$, $\rho_{yx}(0)$ deviates quickly from $h/e^2$, and $\rho_{xx}(0)$ correspondingly increases for $w \leq 1$ μm. In other words, the $\rho_{yx}(0)$ plateau becomes narrower and the $\rho_{xx}(0)$ dip becomes sharper for $w \leq 1$ μm (Figs. 3a to 3f).

We next discuss two possible origins for the confinement-induced QAH decaying behavior in macroscopic Hall bar devices with $w \leq 1$ μm. Both are based on the band structures of the QAH sample, in which the magnetic exchange gap is close to the maximum of the bulk valence bands along Γ-M direction but far from the bulk conduction bands (Fig. 5f) [2,5,32,34]. For $V_g < V_g^0$, the carriers from bulk valence bands are dominant and thus induce the similar gate ($V_g$-$V_g^0$) dependence of $\rho_{yx}(0)$ and $\rho_{xx}(0)$ behaviors in all QAH samples. However, for $V_g > V_g^0$, the CEC and 2D helical surface states on top and bottom surfaces coexist (Fig. 5f) [5]. First, the presence of the helical surface state channels may increase the CEC penetration depth, which favors the occurrence of the confinement-induced interaction between two CECs. Second, the defects and/or disorders inevitably exist and thus induce a few 2D puddles in the bulk state of the QAH samples [18,24,26]. These 2D puddles may extend to larger areas for $V_g > V_g^0$, which assist the occurrence of confinement-induced interaction between two opposite CECs (Fig. 5e).

To further understand the confinement-induced interaction between two opposite CECs, we plot $\rho_{yx}(0)$ and $\rho_{xx}(0)$ at $V_g = V_g^0$ and ($V_g$ - $V_g^0$) = +40 V as a function of $w$. For $V_g = V_g^0$, we find a slow decrease in $\rho_{yx}(0)$ for $w \leq 1$ μm, accompanied by a slow increase in $\rho_{xx}(0)$ (Figs. 4a, 4b, and S5a). However, for $V_g > V_g^0$, both $\rho_{yx}(0)$ and $\rho_{xx}(0)$ decrease much faster for $w \leq 1$ μm (Figs. S5b). We further investigate the current-induced breakdown in these QAH Hall bar devices with 72 nm $\leq w \leq$ 10 μm at $V_g = V_g^0$. Since these QAH Hall bar devices have the same aspect ratio but different lengths $l$, we focus on the current density-induced QAH breakdown and plot the longitudinal electric field $E_x = V_x/l$ as a function of the dc excitation current $I_{dc}$ ($V_x$ is the longitudinal voltage, Fig. 4b). We observe the nonlinear behavior in $E_x$-$I_{dc}$ curves becomes more pronounced in QAH Hall bar devices with smaller $w$, consistent with the prior studies [19,20,24]. A characteristic current $I_0$ is defined as the horizontal intercept of the line fitted in the linear region in $E_x$-$I_{dc}$ curves, which can be used to evaluate the breakdown effect. For the $w \geq 1$ μm samples, $I_0$ is proportional to the Hall bar width $w$ (Fig. 4c), consistent with the current-induced breakdown phenomena in micrometer-size QAH insulators [2,19,20,24]. Since the Hall electric field $E_y \sim h/e^2 \cdot I_{dc}$ increases as $I_{dc}$ increases, our observation indicates that the current-induced breakdown in QAH Hall bar devices with $w \geq 1$ μm is relevant to the Hall electric field [20,24]. However, for the $w < 1$ μm samples, $I_0$ shows a sudden drop (Fig. 4c), which is absent in prior studies [2,19,20,24]. The current-induced QAH breakdown in this regime might be related to the confinement-induced interaction between two CECs, as discussed above.

To investigate the spatial distribution of CECs, we perform theoretical calculations using a thin-film model [10,29,35] and the recursive Green's function approach [36-38]. We choose the magnetic exchange gap of ~3 meV [31], consistent with its Curie temperature $T_C$ ~17 K (Figs. S6 and S7) [2-10,17]. We consider a slab configuration with a periodic boundary condition along the $x$-direction and an open boundary condition along the $y$-direction and plot the local density of state (DOS) as a function of the distance $y$ from one to the opposite edges of the QAH sample at chemical potential μ =0 meV and μ =4 meV with and without the disorder (Figs. 5a to 5d). We find two different regimes for the decaying behaviors of CECs. For μ =0 meV without disorder (i.e. the value of the disorder strength $V_0$ =0 eV), the CEC decays exponentially into the bulk with an intrinsic penetration depth $d$ ~8 nm (Figs. 5a), thus supporting the well-quantized QAH effect in ~72 nm QAH samples (Figs. 1c and 1d). When the disorder is introduced, the CEC first decays exponentially, the same as the clean limit. However, instead of vanishing in the bulk, the CEC retains a long tail of nonzero residual DOS which extends almost as a constant into the bulk region (Fig. 5b). This residual DOS within the magnetic exchange gap can mediate the interaction between two opposite CECs, thus providing an explanation of the experimentally observed slow decaying behavior of the QAH state for 72 nm ≤ $w$ ≤ 1 μm for $V_g = V_g^0$ (Figs. 4a and 4b). We next discuss μ =4 meV, in which the 2D bulk states (from helical surface states of TI films) appear for both clean and disordered cases (Figs. 5c and 5d) and manifest themselves as the residual DOS that hybridize strongly with 1D CECs. Therefore, even though the intrinsic penetration depth of the CEC is short, disorder-induced bulk-like DOS can facilitate the interactions between the two opposite CECs (Fig. 5e) and result in a slow deviation from the perfect QAH state by narrowing the width of the QAH Hall bar devices, as observed in our experiments (Figs. 4a and 4b).

We next study the localization length $\xi$, which characterizes the propagation length of bulk states and thus is expected to control the slow decaying behavior of the QAH state. Figure 5g shows the localization length $\xi$ as a function of $\mu$ at different disorder strength $V_0$. We distinguish two regimes of QAH state. When the chemical potential crosses the 2D bulk bands (i.e.|$\mu$|>1.5 meV) (Fig. 5f), $\xi$ decreases with $V_0$, indicating that the bulk carriers are easier to be localized as the disorder becomes stronger. However, when the chemical potential is located in the magnetic exchange gap (i.e. $|\mu|$ <1.5 meV), $\xi$ increases with $V_0$, as expected that stronger disorders introduce more bulk-like DOS within the magnetic exchange gap, and for $V_0 = 0$, the value of $\xi$ goes towards zero since there are no bulk states. At charge neutral point $\mu =0$ meV, $\xi$ is 200~400 nm depending on the disorder strength $V_0$, consistent with our experimental observation that the two CECs start to interact when the width is smaller than 1 μm (Figs. 4a and 4b). Therefore, by comparing our experiment and theory, we conclude the slow decaying behavior of CECs at the charge neutral point ($V_g = V_g^0$) in experiments results from the hybridization between two CECs mediated by the disorder-induced bulk-like states. $\xi$ continuously increases with $\mu$, which indicates that the QAH state deviates faster when $\mu$ is tuned away from the charge neutral point. This agrees well with our gate-dependent results (Fig. 3), where $\rho_{yx}(0)$ and $\rho_{xx}(0)$ in smaller Hall bar devices deviate from the well-quantized QAH effect much faster for $V_g > V_g^0$ (Fig. S5b).

To summarize, by systematically narrowing the width of the QAH Hall bar device down to ~72 nm, we find that the confinement-induced interaction between two opposite CECs starts to appear in QAH samples with a width of less than 1 μm. The QAH state is found to persist in a Hall bar device with a width of ~72 nm, indicating the CEC width is less than ~36 nm in QAH insulators. We also find that the current-induced QAH breakdown shows a sudden drop at the charge neutral point and the QAH effect decays much faster in the electron-doped regime by reducing the Hall bar width $w$, both of which support the above confinement-induced CEC interaction induced by disorder-induced bulk states in QAH samples. Our work lays down the dimension limitations for the QAH insulators in energy-efficient electronic and spintronic devices. The technique of patterning the quasi-1D QAH structures developed in this work also enables the development of scalable topological quantum computations [2,25,28].

**Acknowledgments:** We are grateful to Y. T. Cui, N. Samarth, and X. D. Xu for helpful discussions. This work is primarily supported by the NSF-CAREER award (DMR-1847811), including sample synthesis, device fabrication, and dilution transport measurements. The PPMS transport measurements are supported by the AFOSR grant (FA9550-21-1-0177). C. Z. C. acknowledges the support from Gordon and Betty Moore Foundation's EPiQS Initiative (GBMF9063 to C. Z. C.).

**Figures and figure captions:**

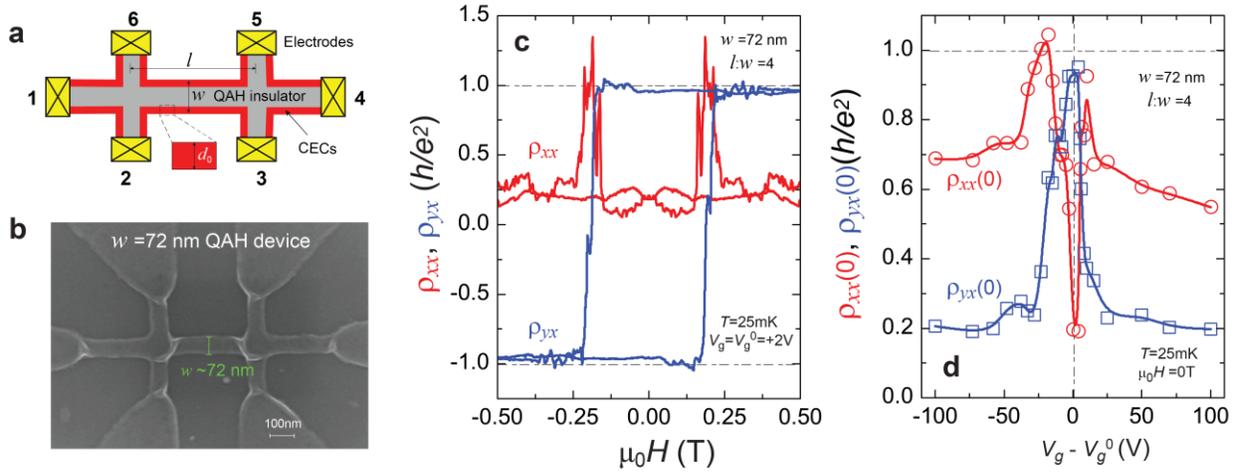

**Fig. 1| The QAH state in a Hall bar device with a width of ~72 nm. a**, Schematic of CECs in a QAH insulator when the current flows through from electrode 1 to electrode 4. **b,** The scanning electron microscope image. **c,** $\mu_0 H$ dependence of $\rho_{xx}$ (red) and $\rho_{yx}$ (blue) measured at $V_g = V_g^0 = +2$ V and $T = 25$ mK. **d**, Gate ($V_g$-$V_g^0$) dependence of $\rho_{xx}(0)$ (red circles) and $\rho_{yx}(0)$ (blue squares) measured at $\mu_0 H = 0$ T and $T = 25$ mK.

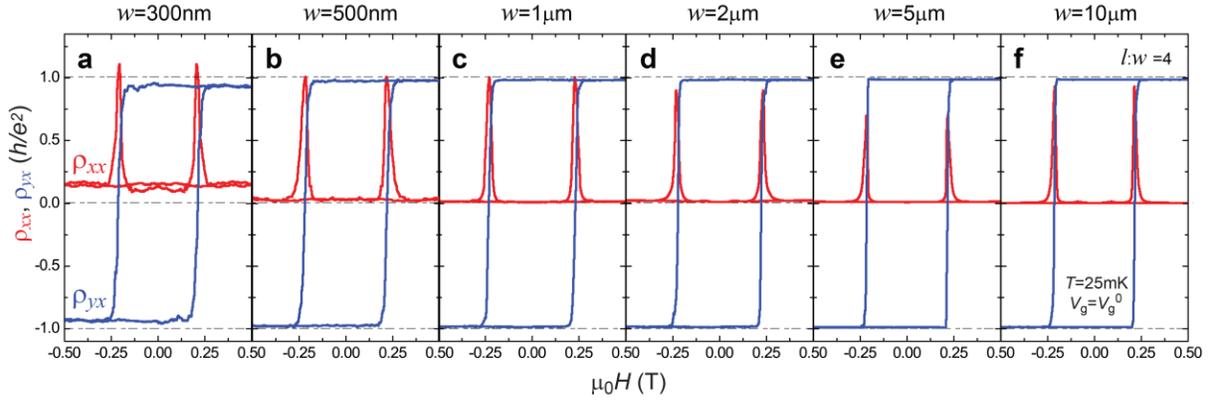

**Fig. 2| The QAH state in Hall bar devices with 300 nm ≤ $w$ ≤ 10 μm at $V_g=V_g^0$. a-f,** $\mu_0 H$ dependence of $\rho_{xx}$ (red) and $\rho_{yx}$ (blue) of the QAH Hall bar devices with $w$ =300 nm (a), $w$ =500 nm (b), $w$ =1 μm (c), $w$ =2 μm (d), $w$ =5 μm (e), and $w$ =10 μm (f). All these measurements are taken at $V_g = V_g^0$ and $T$ =25 mK. The values of $V_g^0$ are +4 V, +6 V, +8 V, +8 V, +11 V, +12 V for the $w$ =300 nm, 500 nm, 1 μm, 2 μm, 5 μm, and 10 μm devices, respectively. All these devices are fabricated by electron-beam lithography.

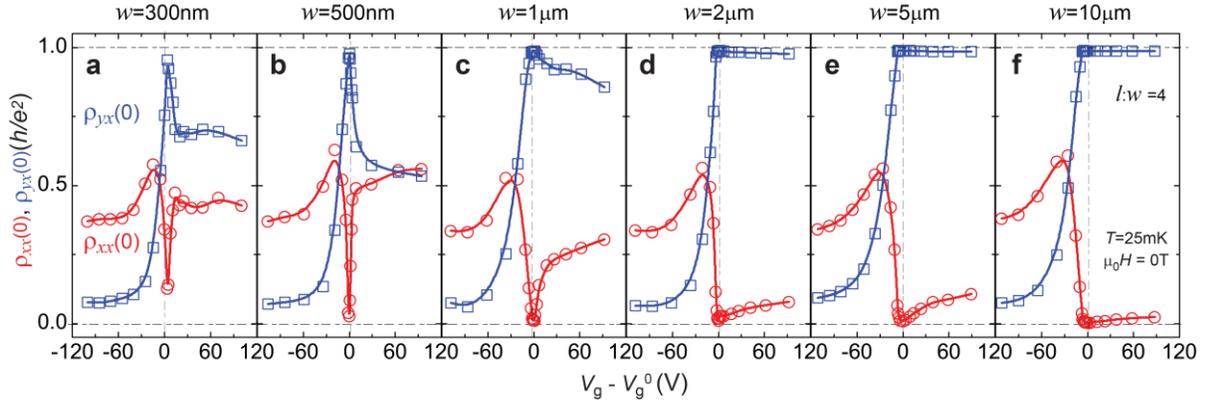

**Fig. 3| Evolution of the QAH state in Hall bar devices with 300 nm ≤ $w$ ≤ 10 μm. a-f,** Gate ($V_g$-$V_g^0$) dependence of $\rho_{xx}(0)$ (red circles) and $\rho_{yx}(0)$ (blue squares) of the QAH Hall bar devices with $w$ =300 nm (a), $w$ =500 nm (b), $w$ =1 μm (c), $w$ =2 μm (d), $w$ =5 μm (e), and $w$ =10 μm (f). All measurements are taken at $\mu_0H$ =0 T and $T$ =25 mK after magnetic training.

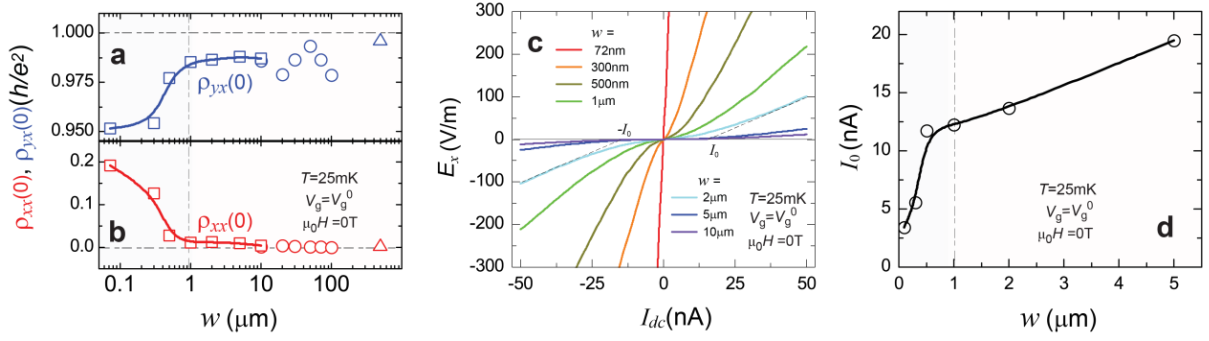

**Fig. 4| Confinement-induced CEC interaction in QAH insulators at $V_g = V_g^0$. a, b,** $w$ dependence of $\rho_{yx}(0)$ (blue, a) and $\rho_{xx}(0)$ (red, b). The square, circle, and triangle points are from the QAH Hall bar devices fabricated by electron-beam lithography, photo-lithography, and mechanical scratching, respectively. **c,** The longitudinal electric field $E_x$ as a function of the dc excitation current $I_{dc}$ for the QAH Hall bar devices with $w \leq 10$ μm. **d,** The characteristic current $I_0$ as a function of $w$.

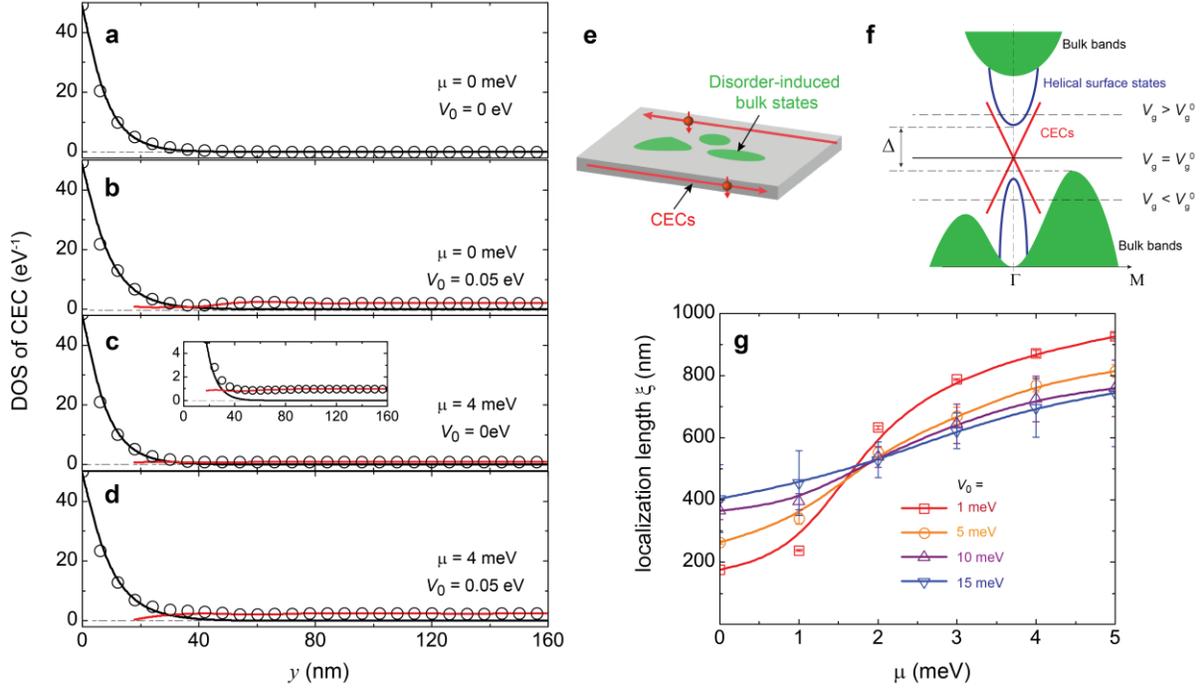

**Fig. 5| Theoretical calculations of confinement-induced CEC interaction in QAH insulators.**
**a-d**, DOS of CEC in a QAH insulator as a function of the depth $y$ at $\mu = 0$ meV and disorder strength $V_0 = 0$ eV (a), $\mu = 0$ meV and $V_0 = 0.05$ eV (b), $\mu = 4$ meV and $V_0 = 0$ eV (c), and $\mu = 4$ meV and $V_0 = 0.05$ eV (d). Inset of (c): enlarged vertical axis range to show the non-zero residual DOS caused by bulk states. **e,** Schematic of CEC interaction induced by the disorder-induced bulk states. **f,** The schematic band structure of the QAH sample. The two CECs (red) do not directly couple to each other, but the bulk states (green) and disorder can mediate the CEC interaction and leads to a slow deviation from the well-quantized QAH effect. **g,** The localization length $\xi$ as a function of $\mu$ at $V_0 = 1$ meV, 5 meV, 10 meV, and 15 meV, respectively. Within the bulk gap ($|\mu| < 1.5 meV$), $\xi$ increases with $V_0$, while in the bulk ($|\mu| > 1.5 meV$), $\xi$ decreases with $V_0$.